\documentclass[aps,twocolumn,prb]{revtex4}
\usepackage{amsfonts}
\usepackage[final,hiresbb]{graphicx}
\usepackage{amsmath}
\usepackage{amssymb}
\usepackage{amstext}

\usepackage[dvips]{color}

\begin{document}

\title{Graphene Nanoengineering and the Inverse-Stone-Thrower-Wales Defect}
\author{Mark T. Lusk$^*$, David T. Wu$^\dag$ and Lincoln D. Carr$^*$}
\affiliation{$^*$Department of Physics, $^\dag$Department of Chemistry and $^\dag$Department of Chemical Engineering,
Colorado School of Mines, Golden, CO 80401, USA}

\begin{abstract}
We analyze a new fundamental building block for monolithic nanoengineering on graphene: the Inverse-Stone-Thrower-Wales (ISTW) defect.  The ISTW is formed from a pair of joined pentagonal carbon rings placed between a pair of heptagonal rings; the well-known Stone-Thrower-Wales (STW) defect is the same arrangement, but with the heptagonal rather than pentagonal rings joined.  When removed and passivated with hydrogen, the structure constitutes a new molecule, \emph{diazulene}, which may be viewed as the result of an ad-dimer defect on anthracene.
Embedding diazulene in the honeycomb lattice, we study the effect of ad-dimers on planar graphene.  Because the ISTW defect has yet to be experimentally identified, we examine several synthesis routes and find one for which the barrier is only slightly higher than that associated with adatom hopping on graphene.  ISTW and STW defects may be viewed as fundamental building blocks for monolithic structures on graphene. We show how to construct extended defect domains on the surface of graphene in the form of blisters, bubbles, and ridges on a length scale as small as 2 \AA~$\times$ 7 \AA.  Our primary tool in these studies is density functional theory.
\end{abstract}

\maketitle


\section{Introduction}

\begin{figure}[ptb]\begin{center}
\includegraphics[width=0.45\textwidth]{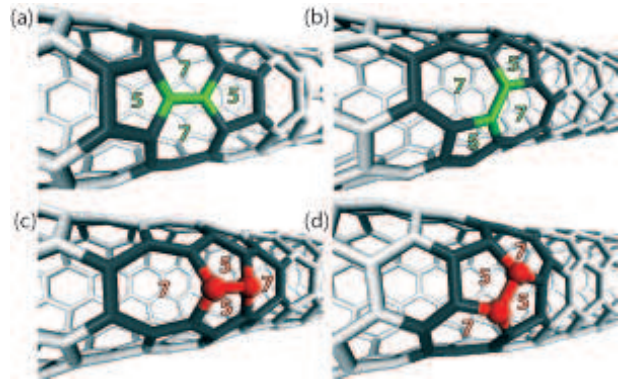}
\caption{(a,b) Stone-Thrower-Wales and (c,d) ad-dimer defects on carbon nanotubes. The defects are shown with achiral (a,c) and chiral (b,d) CNTs.}
\label{CNT}
\end{center}\end{figure}

A substantial fraction of nanotechnology research has focused on the uses of carbon in diamond, graphitic and amorphous forms. Within this setting, defect-based nanoengineering of carbon structures has tended to focus on tubules in order to further extend the technological reach of this novel form of matter.  The removal or addition of carbon atoms to nanotubes has a dramatic impact on their electrical~\cite{lee2005, biercuk2004} and mechanical properties~\cite{salvetat1999, haskins2007} and can be used to weld them together\cite{terrones2002} or break them apart~\cite{ajayan1998}.  These defects can be introduced by ion irradiation~\cite{krasheninnikov2002}, electron irradiation~\cite{smith2001}, and scanning tunneling microscopy (STM)~\cite{berthe2007}.  Beyond single vacancies and single adatoms, defect structures can be introduced which involve more substantial rearrangements of the carbon lattice.  For example, the Stone-Thrower-Wales (STW)~\cite{thrower, stone_1986} defect shown on a carbon nanotube in Fig.~\ref{CNT}(a,b), is constructed by a simple rotation of a pair of carbon atoms to form pairs of opposing 5 and 7-membered rings.

A second class of defects are those which result from the addition of combinations of carbon atoms to the graphene lattice. A properly placed carbon ad-dimer, for instance, results in a different arrangement of  5- and 7-membered rings as shown for carbon nanotubes (CNTs) in  Fig.~\ref{CNT}(c,d). Such ad-dimer defects were first proposed by Orlikowski {\it et al.}~\cite{orlikowski1999} as a means of creating quantum dots from the nanotube constrictions that result from subsequent axial strain. While not yet created in the laboratory, ad-dimer defects have continued to receive attention in theoretical nanotube investigations~\cite{nardelli2000, ewels2002, sternberg2006}. However, the last several years has resulted in significant improvements in the fidelity with which carbon atoms can be manipulated.  STM~\cite{berthe2007} and atomic force microscopy (AFM)~\cite{sugimoto2005} are particularly promising in this regard. A third theoretical possibility is a device that has been designed with the sole purpose of delivering carbon dimers to graphene- and diamond-like materials~\cite{allis2005}.

%
The wealth of knowledge that has been developed for CNTs is of immediate use in considering the entirely new frontier of nanoengineering on graphene, which has only recently been synthesized~\cite{novoselov2005, zhangYB2005}. These single-layer carbon structures have been the subject of significant experimental and theoretical inquiries because of their potential within the electronics industry and because they represent a new and poorly understand form of matter. Just like their tubule counterparts, graphene sheets are known to have defects~\cite{hashimoto_2004, stroscio2007a, stroscio2007b}. At the most basic level, defects can be formed by knocking atoms out of the lattice so as to have vacancies of various sizes; alternatively, additional atoms can be found as adatoms on the surface.  Structural rearrangements within the lattice are also possible, most notably the STW defect, which has an analogous form in graphene as in CNTs. As with nanotubes, ad-dimer defects can be introduced into graphene, as we demonstrated in recent work~\cite{LuskCarr2008a}.  Within this setting, the defects appear as smooth bumps protruding out of the graphene plane.  Because the ad-dimer defect has adjacent 5-membered rings instead of adjacent 7-membered rings, it was dubbed an \textit{Inverse Stone-Thrower-Wales (ISTW) defect}. Taken together, the STW and ISTW defects may be viewed as basic building blocks in a taxonomy of structures with novel thermo-mechanical, chemical, electrical, and magnetic properties.

The study of this defect taxonomy is necessarily distinct from the previous consideration of nanotubes with similar defects~\cite{orlikowski1999,nardelli2000,ewels2002,sternberg2006}. There the emphasis has been on the spreading of defects in response to axial strain. The planar geometry of graphene offers a simpler setting for developing a fundamental understanding of how to build stable structures on unstrained graphene. The planar setting also leads to a consideration of patterned defects and facilitates a connection between graphene and other planar carbon allotropes, as we have demonstrated elsewhere~\cite{LuskCarr2008b}.

The present work provides a more detailed structural and synthesis analysis of the ISTW defect than was provided in our initial presentation~\cite{LuskCarr2008a}.  Density functional theory is employed to determine the ground state of an isolated defect within both periodic and freestanding passivated settings. The associated formation and binding energies are compared to assess the range over which the structures influence lattice strain. The synthesis techniques originally considered in carbon nanotubes apply to the graphene setting as well.  In addition, four synthesis routes are introduced and the simplest, coalescence of two adatoms, is examined in detail.  ISTW defects are then aligned on open and closed contours to illustrate the potential for making technologically important carbon structures using this single building block.  Such structures may serve as quantum wires and quantum dots for quantum-coherent graphene-based electronics.  Linear arrangements of ISTW defects result in stable ridges while closed contours put a compressive stress on the internal patch of graphene, causing it to raise up.

Through the introduction of an STW defect, ISTW defects can be partially dissociated. This is reminiscent of the partial dissociation of dislocations that occurs in three-dimensional crystals, but here there are two dislocations that are pulled apart by the presence of a third. Since a pair of 5/7 rings corresponds to a dislocation, the ISTW defect may be viewed as two tightly coupled dislocations. The introduction of another dislocation--i.e. a STW defect--causes the ISTW dislocations to partially separate. The resulting {\em ISTW-STW blister} is still out-of-plane but has three-fold symmetry and a larger footprint than the original defect. For the sake of clarity in the present work, the term \emph{blister} refers exclusively to such structures. Blisters exhibit three-fold symmetry and can be used as building blocks for even more complex carbon structures. In contrast to a strain induced bond rotation which results in blisters on CNTs~\cite{orlikowski1999}, we examine the notion that STW defects may actually serve to catalyze the creation of ISTW defects by attracting adatoms.

This article proceeds to elucidate defect nano-engineering as follows.  In Sec.~\ref{sec:theory}, we present a brief review of density functional theory (DFT) and our use thereof.  In Sec.~\ref{sec:ISTW}, we apply DFT to the ISTW defect and show how this defect creates a blister in the graphene plane. In Sec.~\ref{sec:synthesis} we discuss several synthesis routes to the creation of this defect, focusing on a new method which is complementary to our original proposal~\cite{LuskCarr2008a}.  In Sec.~\ref{sec:structures} we describe extended nano-structures, including embedded ridges bubbles of graphene corralled from the parent material by a closed defect ridge.  Finally, in Sec.~\ref{sec:conclusion}, we conclude.

\section{Theoretical Methods}
\label{sec:theory}

Our analyses seek first to identify stable defect structures in the absence of finite temperature effects. An adiabatic approximation is therefore made which assumes that electrons are in their ground state. Nuclei are treated as classical point masses, and are taken to have fixed positions during a calculation of the electronic ground states, i.e., we make a Born-Oppenheimer approximation. Given the electron density, the Hellman-Feynman theorem~\cite{feynman1939} is then used to determine the forces on the nuclei, whose positions are then incrementally adjusted. This procedure of calculating electron density followed by ion nudging is repeated until the nuclei themselves are in local equilibrium. The system is then referred to as being geometrically optimized. The key is therefore to determine the ground state electron density, and this is accomplished using Density Functional Theory (DFT)~\cite{parr1989}. 

Correlations between electrons and the enforcement of wave function antisymmetry are approximated with a single \textit{exchange-correlation} energy functional. We employ an approximation in which nonlocal effects are approximated through a dependence on the gradient in electron density. Within this \textit{generalized gradient approximation} (GGA), we use the Perdew-Wang parameterization.  This involves no empirical parameters and uses the uniform electron gas as a basis~\cite{perdew1992}.  

The Kohn-Sham orbitals used in all calculations are atom-centered functions constructed with a radial component that is numerically calculated.  For each carbon atom, 1$s$, 2$s$, 2$p$, and 3$d$ orbitals were supplemented with 2$s$ and 2$p$ orbitals obtained from the $C^{+2}$ ion\cite{delley1990}. For periodic calculations, this \textit{double-numeric, polarized} basis set was placed within a Bloch function setting so that the orbitals used, $\phi_i$, included a plane wave component. The number of basis functions therefore scaled linearly with the number of discrete points used to sample the Brillouin zone. Periodic boundary conditions were employed and vacuum slabs were used to isolate the replicated graphene layers.  This approach to geometric optimization was carried out using the DFT code, DMOL~\cite{delley1990}.  

As a check on the physical validity of the method, the ground state energy of C$_{60}$, i.e., a buckyball, was estimated to be 384 meV/atom above that of graphene, consistent with a literature value of 380 meV/atom~\cite{terrones2004}. Likewise, a single Stone-Thrower-Wales  defect was estimated to have a formation energy of 5.08 eV when embedded within a 144-atom graphene supercell; this compares well with a separate DFT estimate of 4.8 eV from the literature~\cite{li2005}.

To explore the barriers to carbon re-structuring, a hybrid linear synchronous transit/quadratic synchronous transit (LST/QST) transition state search algorithm was employed~\cite{govind2003}.  The LST algorithm finds a transition state by interpolating geometrically between a reactant and product whose corresponding coordinates are at maximum coincidence. The energy is evaluated for each element of this interpolation set. The set with maximum energy is then geometrically optimized using a conjugate gradient minimum search. This configuration is then used in a QST search step which is a three-point-interpolation. The new pathway is defined jointly by the newly found structure and the two path-limiting structures. The LST/QST process is repeated until iterations on the transition state converge. At the conclusion of a successful transition state calculation, a vibrational analysis is performed to confirm that the transition configuration is stationary. A true transition state will have one imaginary vibrational frequency whose normal mode corresponds to the reaction coordinate while all other frequencies will be real. As a check on the method, a barrier of 0.52 eV was obtained for adatom hopping between adjacent bridge sites on graphene. This is consistent with a DFT estimate of 0.45 eV obtained elsewhere~\cite{krasheninnikov2004}.

The DFT method, as implemented in DMOL, has been compared to experimental results for a wide range of simple systems~\cite{delley2000b} and has been subjected to a rigorous numerical analysis~\cite{delley1990, delley2000}. As previously noted, we have included comparisons of our results, both with experiments and other DFT calculations.  In addition, we performed a convergence study in order to measure the numerical uncertainty associated with our results.  To this end, a single, primitive two-atom graphene cell was scrutinized with the number of k-points, cut-off radius, and integration accuracy analyzed.  The convergence study indicates that a use of cutoff radius of 3.5\AA, a k-point density equivalent to 40 k-points for a graphene cell, and a ``medium'' integrator setting will produce numerical results which are accurate to three significant digits.  With these settings, the primitive cell side length and carbon bond length were found to be within 0.01\% of experimentally obtained values\cite{trucano1975}. These are the settings used in all of the results presented in this work.

\section{The Inverse Stone-Thrower-Wales Defect}
\label{sec:ISTW}

\begin{figure}[ptb]\begin{center}
\includegraphics[width=0.45\textwidth]{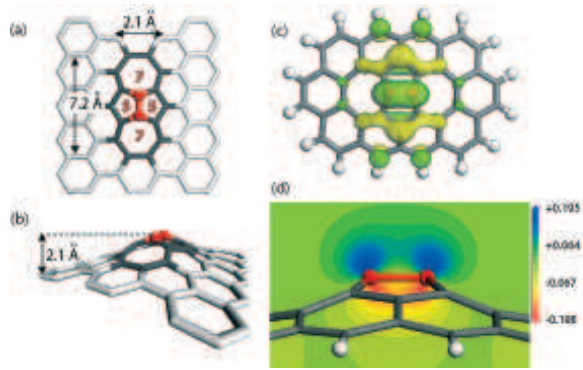}
\caption{(a, b) Top and perspective views of an ISTW defect. (c,d) Pi-bonding is shown as lobes above and below a hydrogen-passivated ISTW defect: (c) iso-surface (0.03) of electron density for the highest occupied molecular orbital; (d) electron density cross section, centered on the adatoms, of the highest occupied molecular orbital. The units of both iso-surface and cross-section are electrons/\AA$^3$.}
\label{ISTW_pi_bonding}
\end{center}\end{figure}

We now turn to the consideration of ISTW defects using DFT. A single ad-dimer was placed above opposing bridge sites on a hexagonal ring of carbon atoms. When geometrically optimized, these adatoms form a backbone for adjacent 5-membered rings as shown in figure Fig.~\ref{ISTW_pi_bonding}(a, b).  The defect involves the re-structuring of three hexagonal rings laid out in a row, so that the final ISTW-defect structure is elongated, as can be observed in the figure.  All carbon atoms still covalently bond with three nearest neighbors and so have unbonded $\pi_z$ electrons out of plane which give the graphene much of its interesting charge transport character.  This is clear, for instance, in the density cross-section for the highest molecular orbital of a hydrogen-passivated ISTW defect shown in Fig.~\ref{ISTW_pi_bonding}(c, d). However, atoms associated with the ISTW defect exhibit bond distortion that may result in localized electrical and magnetic anomalies.

\begin{figure}[ptb]\begin{center}
\includegraphics[width=0.45\textwidth]{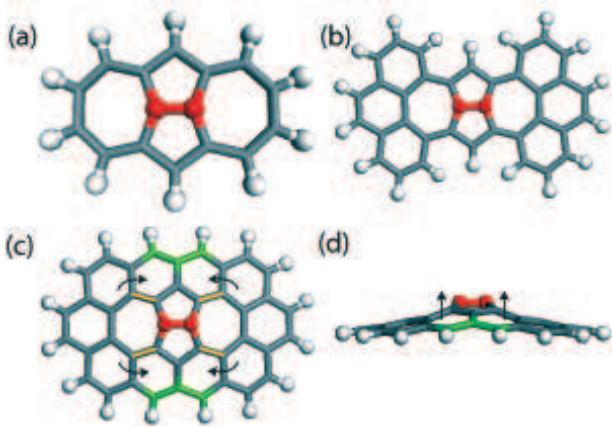}
\caption{Geometric explanation for the out-of-plane distortion of an ISTW defect. (a) ISTW defect without graphene has planar ground state and constitutes a new molecule, (b) addition of 6 of the 10 carbon rings required to encircle the ISTW defect also has a planar ground state, (c) addition of the remaining carbon rings pulls the left and right sides together at both top and bottom, and this causes out-of-plane distortion, (d) side-view of image in (c).}
\label{ISTW_buildup}
\end{center}\end{figure}

The ground state structure of ISTW defects is non-planar, and one is led to consider what causes the distortion out of plane.  After all, there are a number of planar allotropes of carbon that feature 5- and 7-membered rings, and the hydrogen-passivated ISTW defect by itself is planar, as shown in Fig.~\ref{ISTW_buildup}(a).  In fact, this object constitutes a new C$_{16}$H$_{10}$ molecule.  Its ground state energy is 2.79 eV higher than the more common benzoidal pyrene, but is 0.07 eV lower in energy than the previously identified pentalenoheptalene~\cite{birss1971}, composed of two fused azulene structures which can also be viewed as fused pentalene and heptalene. Because it does not include a heptalene moiety and is the most symmetric result of combining two azulenes, we dub the new molecule \emph{diazulene}.

Panels (b)-(d) of Fig.~\ref{ISTW_buildup} provide a simple geometric explanation for the out-of-plane distortion of an ISTW defect embedded in graphene.  The diazulene molecule is surrounded by a periphery of ten carbon hexagons. Six of them do not cause any significant changes in existing bond length (Fig.~\ref{ISTW_buildup}(b)), but the last four hexagons require that four bonds be rotated by several degrees (arrows in Fig.~\ref{ISTW_buildup}(c)), and this restriction to planar deformation would require a significant reduction in the length of the four bonds highlighted in gold in that figure.  An energetically favorable option is for these four bonds to rotate out of plane in order to accommodate the final four hexagons.

\begin{figure}[ptb]\begin{center}
\includegraphics[width=0.45\textwidth]{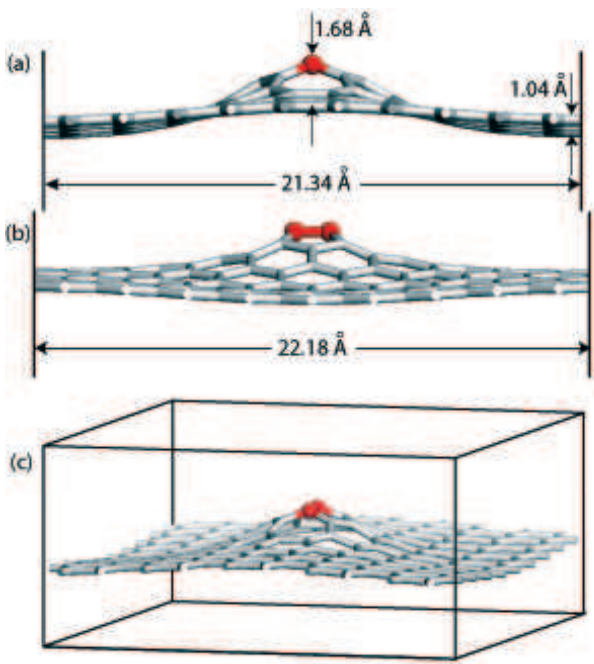}
\caption{ISTW defect on periodic domain with 182 carbon atoms showing (a) edge view in which adatoms are aligned, (b) edge view in transverse to adatom alignment, and (c) perspective view. The black lines indicate the boundaries of the cell.}
\label{ISTW_182}
\end{center}
\end{figure}

When placed within an extended graphene matrix, the ISTW defect causes a distortion of the graphene lattice which extends over tens of angstroms. In order to assess the extent of this influence, and to better estimate the defect formation energy, periodic domains of increasing size were considered. The ground state energy $E_{\mathrm{G144}}$ of a graphene sheet composed of 144 carbon atoms was first calculated.  The energy $E_{\mathrm{ISTW},N}$ of a second graphene sheet with a single ISTW defect and a total of $N$ atoms was then calculated. The formation energy was calculated as
\begin{equation}
E_f = E_{\mathrm{ISTW},N} - E_{\mathrm{G144}}*\frac{N}{144} .
\end{equation}
For $N=$ 100, 162, and 182,  the formation energies are 5.82 eV, 6.21 eV and 6.22 eV, respectively. This suggests that the formation energy is converged within 0.01 eV, the limit of numerical accuracy for our computational approach. As an additional check on convergence, the formation energy for the 182-atom system was calculated with 27 and 8 irreducible k-points instead of the 48 irreducible k-points used to obtain 6.22 eV. The energies are 6.19 eV (28 points) and 6.25 (8 points). Since k-point convergence is typically oscillatory, the data suggests that the formation energy is k-point converged to within 0.03 eV.  Figure~\ref{ISTW_182} shows the ground state of an ISTW defect for the periodic 182-atom system.

In a similar manner, the ISTW binding energy was calculated by treating the defect as an adsorbed $C_2$ dimer using 
\begin{equation}
E_b =(E_{G,N-2} + E_{\mathrm{C_2}}) - E_{\mathrm{ISTW}, N} .
\end{equation}

\begin{figure}[ptb]\begin{center}
\includegraphics[width=0.45\textwidth]{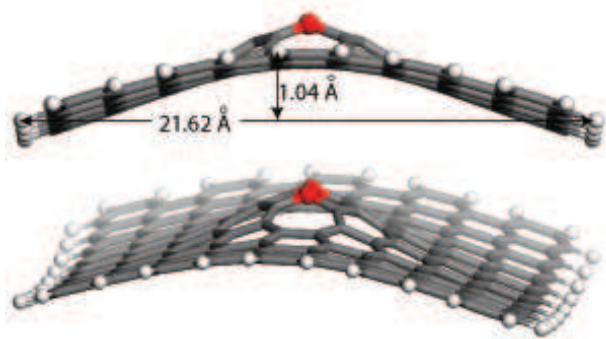}
\caption{Side (a) and perspective (b) views of ISTW defect on a freestanding, hydrogen passivated graphene sheet with 182 carbon atoms.}
\label{ISTW_182_free}
\end{center}\end{figure}

Here $E_{G, N-2}$ is the energy of defect-free graphene, $E_{\mathrm{ISTW, N}}$ is the energy of the same sheet with a single ISTW defect, and $E_{\mathrm{C_2}}$ is the energy of a single carbon dimer.  A positive binding energy indicates that a defect structure is energetically favorable. For $N=$ 100, 162, and 182, the binding energies are 3.31 eV, 2.92 eV and 2.91 eV, respectively. These binding energies in periodic domains can be compared with those associated with free-standing, hydrogen-passivated sheets of graphene. Figure~\ref{ISTW_182_free} shows the resulting ground state structure for a sheet with 182 carbon atoms. The associated binding energy is 3.29 eV. The 0.38 eV discrepancy between free standing and periodic systems reflects the ability of a finite free standing system to distort to relax strain energy, as can be seen in the figure.  This suggests that embedding the defect within an even larger system (periodic or free) would give a more accurate estimate for the binding energy and is an indication that the ISTW defect affects the state of lattice strain out beyond 11\AA.  As a point of comparison, the freestanding, hydrogen-passivated graphene sheet was constrained so that the carbon atoms on its border were fixed in the positions shown in Fig.~\ref{ISTW_182_free}(b). All of the other atoms were allowed to relax. The difference in energy between the ground state obtained and that associated with the planar, passivated graphene sheet of the same size is only 0.65 eV.

\begin{figure}[ptb]\begin{center}
\includegraphics[width=0.45\textwidth]{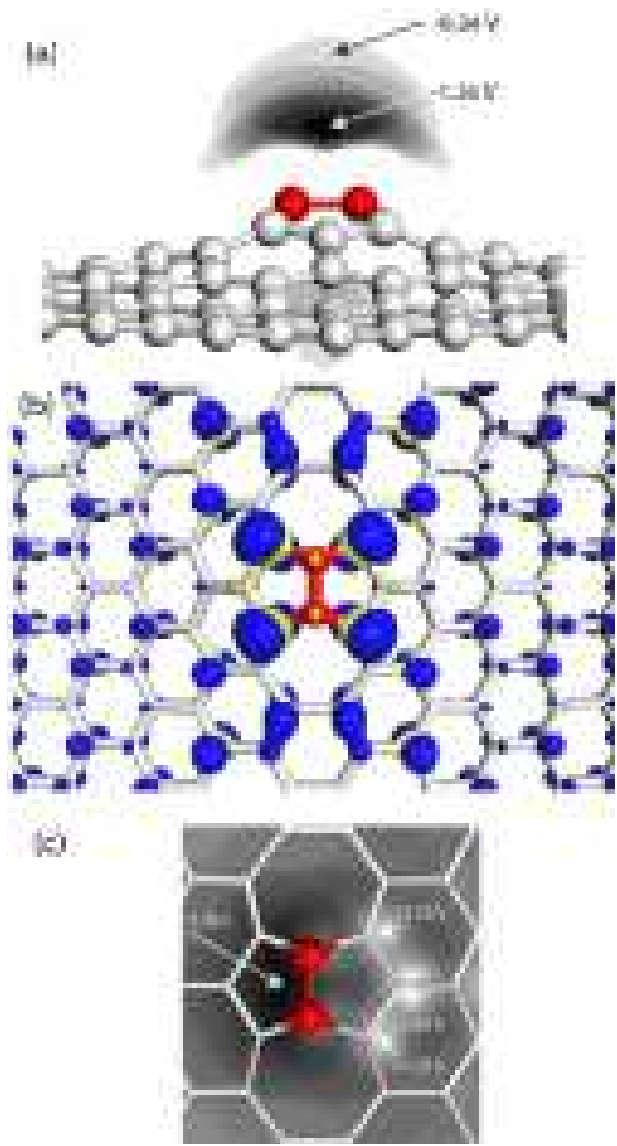}
\caption{ISTW defect on periodic domain with 182 carbon atoms showing (a) electrostatic potential well of 1.36 V atop the defect, (b) electron density isosurface of the difference in electron density between charged and uncharged systems, (Blue indicates a value of 0.002 electrons/$\AA^3$ while yellow is the negative of this. The integral of this density field is equal to unity.), and (c) electrostatic potential in a plane $1.2~\AA$ above the defect tangent plane of its rising right surface. (The perspective is from beneath the defect in order to clearly show the relation of the extrema to the carbon nuclei.)}
\label{ISW182_V_crosssection}
\end{center}\end{figure}

Although our primary focus is on the structural character of defect building blocks, an initial assessment of the charge confinement of ISTW defects has been carried out. The electrostatic potential was first calculated for the configuration shown in Figure~\ref{ISTW_182}. A cross section of the result is plotted as Figure~\ref{ISW182_V_crosssection}(a) which shows that there is a local minimum,  with a depth of 1.36 eV, in the electrostatic potential above the neutral ISTW defect.  Additional electrons will therefore tend to be repelled from the region immediately above the defect.  However, the ISTW defect will still attract additional charge to its shoulders.  This is shown in Figure ~\ref{ISW182_V_crosssection}(b) as a map of the difference in charge density between a charged system and neutral system. The additional electron is localized at the ISTW defect above nuclei which are nearest neighbors to the carbon ad-dimer. Here the electrostatic potential is positive. Figure ~\ref{ISW182_V_crosssection}(c) quantifies this in a slice of the electrostatic potential field parallel to the rising surface of the ISTW defect and $1.2~\AA$ above the defect tangent plane. We conclude that the ISTW defect will attract and confine additional negative charge.

\section{Synthesis of an Inverse Stone-Thrower-Wales Defect}
\label{sec:synthesis}

In principle, the construction of an ISTW defect is a straightforward matter of placing a correctly oriented carbon ad-dimer on the surface of graphene.  The yet-to-be built \textit{DC10c} machine of Allis and Drexler, for instance, is designed to do precisely this~\cite{allis2005}, and there are indications that an STM approach may be more immediately possible~\cite{berthe2007}.  Interestingly, an STM was used to produce {\em domed features} that are 2\AA~high with a footprint diameter of 7\AA~on highly ordered pyrolytic graphite in liquid water~\cite{heben1991}. These dimensions are close to the size of an ISTW defect with a height of 2.1\AA~and a footprint of 12.2\AA~$\times$7.1\AA.  Atomic force microscopy offers a third method which may also provide for direct construction of ISTW defect structures~\cite{sugimoto2005}.  These and other potential methodologies would certainly be aided by the study of existing ISTW defects. Since an ISTW defect has not been conclusively identified in experiments, a reasonable first step would be to determine the likelihood that they exist naturally in graphene structures.  Their formation may be facilitated by other defects in the carbon sheet or it may be that the chance meeting of two adatoms results in an ISTW defect.

\begin{figure}[t]\begin{center}
\includegraphics[width=0.45\textwidth]{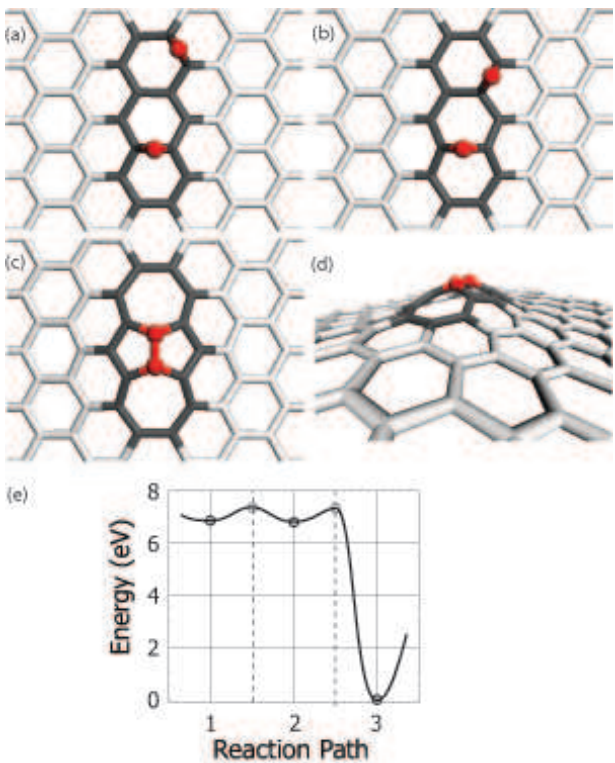}
\caption{Synthesis of an ISTW defect by meeting of two adatoms.  The top adatom takes two hops towards the other via bridge sites from a position (a) in which the adatom interaction is negligible. The hopping barriers are 0.57 eV and 0.44 eV, respectively. The second hop causes a significant re-structuring with a 6.89 eV drop in energy. The result is an ISTW defect (c, d).}
\label{ISTW_synthesis_adatoms}
\end{center}\end{figure}

Motivated by this reasoning, we previously considered the possibility of synthesizing ISTW defects on the lip of divacancies~\cite{LuskCarr2008a}.  It was hypothesized that vacancies facilitate significant low energy distortions of the two-dimensional graphene manifold making it a natural setting for ISTW formation. Adding to the attractiveness of divacancies for ISTW formation, such sites have been experimentally observed to attract adatoms~\cite{hashimoto_2004}. To explore the likelihood that divacancies might facilitate the formation of an ISTW defect by adatom coalescence, an estimate was made of the most likely reaction path for ISTW defect formation on a divacancy. An LST/QST algorithm~\cite{govind2003} was used to construct a reaction pathway for the approach of a single adatom to a divacancy.  Adatoms hop between adjacent bridge sites, and a reaction path was identified by which a single adatom moves to the periphery of the divacancy. The principal reaction barrier and net reaction energy were estimated to be 1.55 eV and -2.60 eV, respectively. Adatoms are therefore likely to be found in such positions since the system can give up 2.6 eV by if an adatom hops to the periphery. This is supported by experimental observation~\cite{hashimoto_2004}.  The likelihood of a second adatom coalescing with such a trapped adatom was then considered. As the second adatom hops towards the di-vacany, the last two hops must overcome an energy barrier much higher than the 0.52 eV barrier found associated with pristine graphene. However, the system relaxes after each of these hops so that the overall system energy decreases after each one. The last hop must overcome the largest barrier, 1.55 eV, to create an ISTW defect. Relative to a configuration in which the second adatom is far removed from the divacancy, the final ground state energy is reduced by 7.1 eV~\cite{LuskCarr2008a}. Energetically speaking, a second adatom wandering near the divacancy will be drawn in to create an ISTW defect because each hop lowers the system energy.

In order to assess the degree to which the divacancy catalyzes the ISTW formation, we now consider adatom coalescence on a perfect graphene lattice, i.e., by the random meeting of two adatoms.  This is shown in Fig.~\ref{ISTW_synthesis_adatoms}(a-d).  Once again, an LST/QST algorithm was used~\cite{govind2003}.  A plot of the energy vs. reaction coordinate for the lowest barrier path is provided in Fig.~\ref{ISTW_synthesis_adatoms}(e).  The associated reaction energies are $E^r_{12} = -0.14$ eV and $E^r_{23} = -6.89$ eV, while the reaction barriers are $E^b_{12} = 0.57$ eV and $E^b_{23} = 0.44$ eV.  Surprisingly, this indicates that the energy gained by ISTW formation is nearly as great as when it forms on the lip of a divacancy. Furthermore, the primary reaction barrier is only one-third as large as that associated with divacancy mediation and is essentially the same as the barrier for simply adatom hopping (0.52 eV). Of course, the divacancy offers the advantage of trapping adatoms on its lip while freely roaming adatoms may not readily find each other. In both scenarios, the large reaction energy is transferred to phonons in the defected graphene.

Since divacancies facilitate the formation of ISTW defects, one may wonder whether single vacancies might play a similar role. This seems contrary to intuition, though, since a single vacancy does not result in significant lattice relaxation. The possibility was considered, though, using the LST/QST algorithm. It was found that a single adatom can approach and occupy a vacancy with no barrier greater than 0.62 eV, only 0.1 eV higher than simple adatom hopping on graphene. ISTW defects are thus unlikely to form on single vacancies because one of the two adatoms is very likely to occupy the vacancy.

\begin{figure}[ptb]\begin{center}
\includegraphics[width=0.45\textwidth]{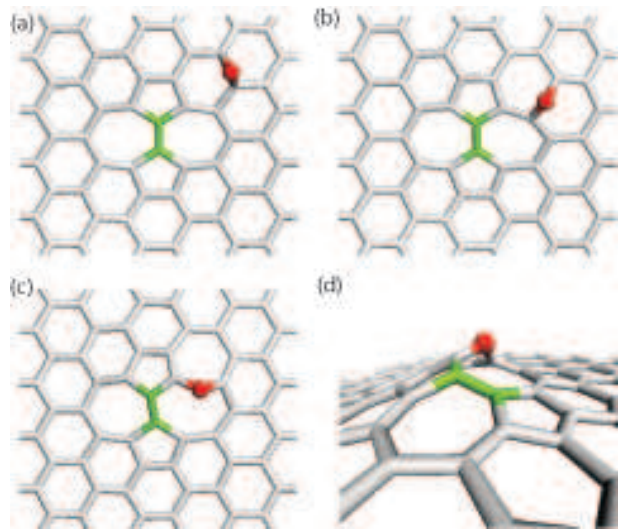}
\caption{The first adatom hops to an STW defect in the synthesis of an ISTW-STW blister:
(a)-(c) correspond to A1 - A3 in Fig.~\ref{blister_synth_rxnpath}(a).  (d) Side perspective
view of (c).}
\label{STW_ISTW_blister_part1}
\end{center}\end{figure}

\begin{figure}[ptb]\begin{center}
\includegraphics[width=0.45\textwidth]{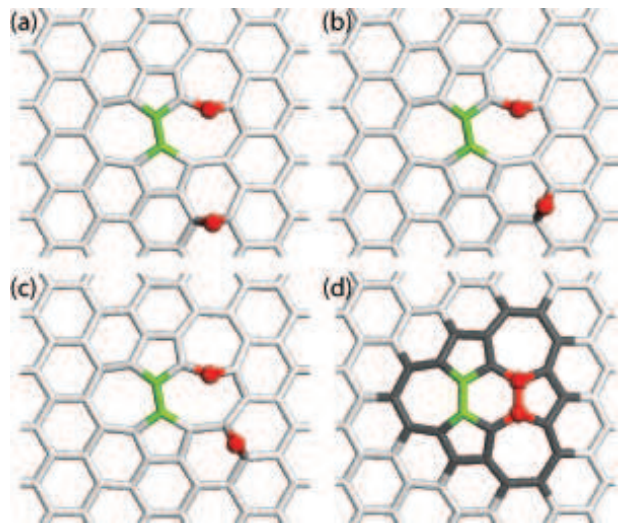}
\caption{The second adatom hops to an STW defect in the synthesis of an ISTW-STW blister, following
from Fig.~\ref{STW_ISTW_blister_part1}:
(a)-(d) correspond to B1 - B4 in Fig.~\ref{blister_synth_rxnpath}(b).}
\label{STW_ISTW_synth_adatom}
\end{center}\end{figure}

\begin{figure}[ptb]\begin{center}
\includegraphics[width=0.45\textwidth]{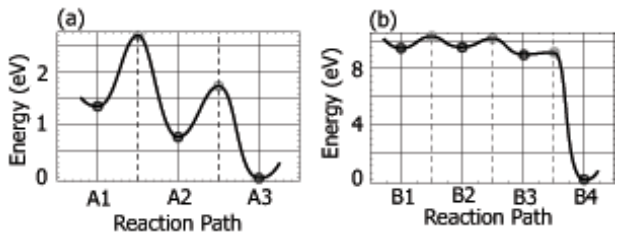}
\caption{Reaction path for creation of an ISTW-STW blister by adding adatoms to an STW defect, corresponding to Figs.~\ref{STW_ISTW_blister_part1}-\ref{STW_ISTW_synth_adatom}.  (a) Path A1-A3 tracks the hopping of the first adatom to the STW defect (Fig.~\ref{STW_ISTW_blister_part1}). (b) Path B1-B4 tracks the hopping of the second adatom (Fig.~\ref{STW_ISTW_synth_adatom}).}
\label{blister_synth_rxnpath}
\end{center}\end{figure}

STW defects are commonly found on graphene, and they may serve as catalytic sites for the formation of an ISTW defect. The attraction of two adatoms to a STW defect would result in a blister (partially dissociated ISTW defect). A representative synthesis path, evaluated using the LST/QST algorithm, is shown in Figs.~\ref{STW_ISTW_blister_part1} and~\ref{STW_ISTW_synth_adatom}. The associated energy plots are given in Fig.~\ref{blister_synth_rxnpath}.  The first adatom must overcome barriers of 1.72 eV and 1.34 eV in order to approach the STW defect with a net energy reduction of 1.35 eV. This is qualitatively different than the situation with a divacancy where the first adatom overcomes barriers of ~0.52 eV until it becomes trapped in the energy well at the divacancy periphery. Specifically, the divacancy will hold the first adatom at a site from which the ISTW defect can be directly formed. In the case of catalysis using a STW defect, the first adatom will tend to be repelled by the defect, by two barriers, before arriving at the periphery. If the first adatom were to arrive at the periphery, though, the second atom energy barriers are relatively low. It must overcome barriers of 0.90 eV, 0.62 eV and 0.13 eV as it approaches with a net energy reduction of 9.47 eV. The system is therefore very stable once formed, but the large double barrier to the first adatom makes the entire synthesis route less plausible than that possible with a divacancy which has a maximum reaction barrier of 1.55 eV. 

Once an ISTW defect forms, a large amount of energy is released and suggests that the reverse process will not occur. However, it may be that a subsequent re-structuring takes place. In particular, a blister may be formed by partially dissociating the ISTW dislocations using an STW defect.  The notion of defect dissociation calls to mind the formation of partial dislocations in crystalline solids~\cite{hirth1968}. Such ISTW dissociations have been theorized to be energetically favorable under sufficiently large strains in carbon nanotubes~\cite{orlikowski1999}.  

The simplest such dissociation occurs when a STW defect separates an ISTW defect into three end-to-end pairs of 5/7 ring pairs, as shown in Fig.~\ref{STW_ISTW_synth_adatom}(d).  This removes back-to-back pentagons at the expense of introducing an additional bond distortion into the lattice. Unlike the ISTW defect, the resulting blister has three-fold symmetry. It involves 24 atoms and stands at a height of 1.9\AA~out of the graphene plane. Within the periodic 182-atom setting previously considered, DFT calculations shows that the formation energy for a blister is 0.46 eV higher than that for a ISTW defect. This indicates that the dissociated state is less likely to be observed unless other influences come into play. In addition, the barrier for this process is 6.23 eV. Taken together, the ISTW defect is not inclined to partially dissociate once formed. 


In addition to the two approaches identified to create ISTW defects, two other possibilities are briefly noted.  Carbon dimers could be propelled towards graphene where they might be trapped as ISTW defects.  These double bonded dimers have a bond length of 1.24\AA~\cite{forni1995}, while the bond length of the ISTW adatoms is 1.4\AA. The primary issues are therefore dimer kinetic energy and orientation.  A preliminary quantum molecular dynamics analysis indicates that an initial kinetic energy of 15.7 eV per carbon atom is sufficient to overcome reaction barriers while not breaking through the graphene; an initial kinetic energy of 7.0 eV per carbon atom resulted in an elastic bounce of the $C_2$ while 35.3 eV per atom caused the dimer to rupture the graphene without attachment.  Dimers with the correct kinetic energy result in an ISTW defect. The carbon dimers can be produced in situ~\cite{goyette1998}. It may also be possible to use acetylene instead of diatomic carbon.


\section{Extended Defect Domains}
\label{sec:structures}

\begin{figure}[ptb]\begin{center}
\includegraphics[width=0.45\textwidth]{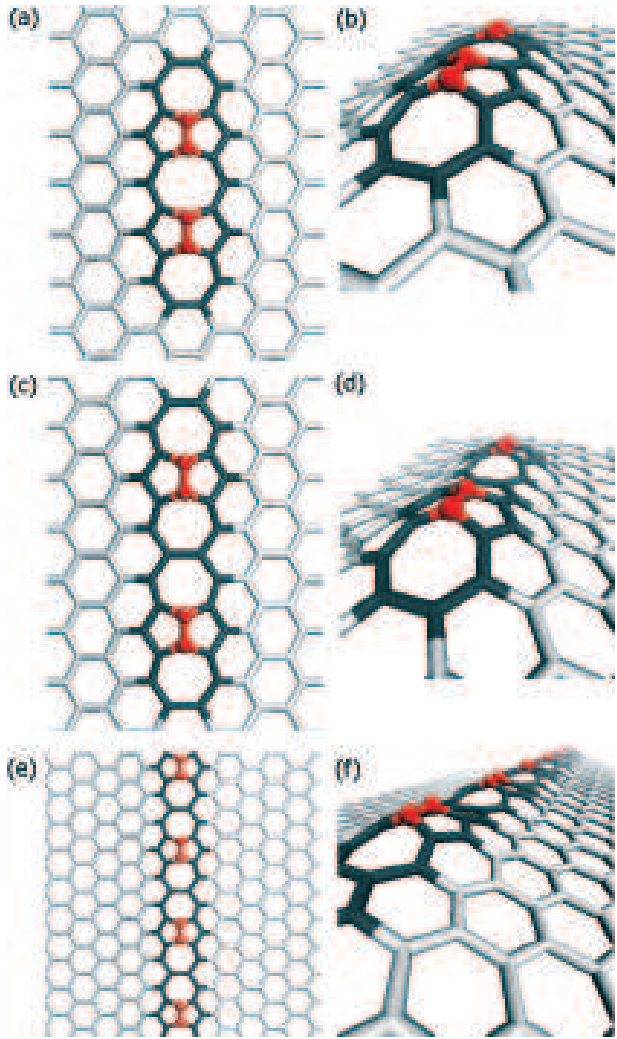}
\caption{Linear arrangements of ISTW defects: (a)-(b) close spacing of two defects; (c)-(d) wider spacing of two defects; (e)-(f) a monolithic ridge of defects that could be used to influence the direction of charge transport as a nano-wire in monolithic graphene-based electronics}.
\label{ISTW_multi_linear}
\end{center}\end{figure}
\begin{figure}[ptb]\begin{center}
\includegraphics[width=0.45\textwidth]{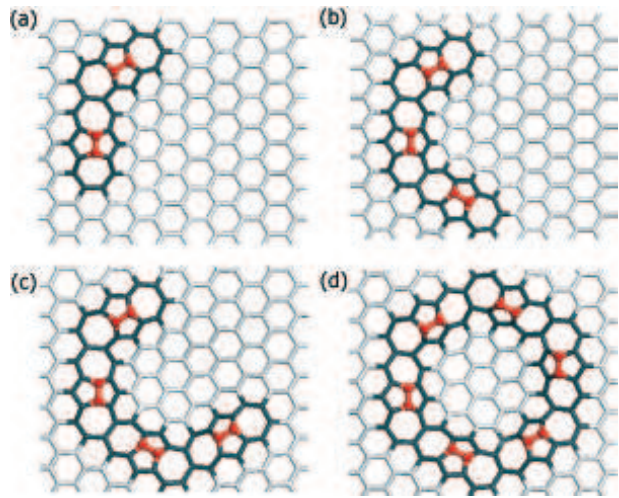}
\caption{Nonlinear arrangements of ISTW defects can be used to create a closed contour which results in a raised graphene bubble. Top view.}
\label{ISTW_multi_nonlin_top}
\end{center}\end{figure}
\begin{figure}[ptb]\begin{center}
\includegraphics[width=0.45\textwidth]{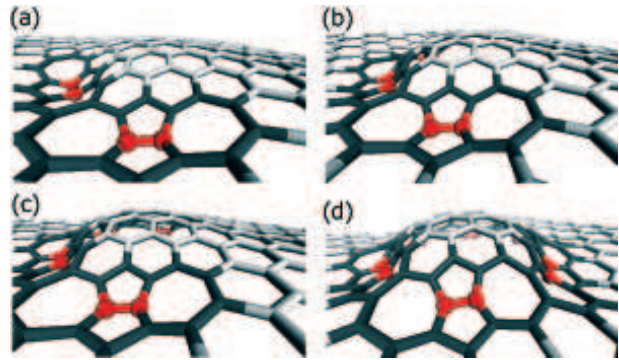}
\caption{Graphene bubble: side perspective view of Fig.~\ref{ISTW_multi_nonlin_top}.}
\label{ISTW_multi_nonlin}
\end{center}\end{figure}

Having elucidated the structural and synthesis features of a single ISTW defect, we now consider them in a variety of extended settings.  A linear arrangement of ISTW defects results is an undulating, raised structure which we refer to as a \textit{ridge}, and two types of such ridges are examined.  Two defects may be overlapped so as to share an octagonal ring as shown in Fig.~\ref{ISTW_multi_linear}(a, b), but a wider spacing, as shown in Fig.~\ref{ISTW_multi_linear}(c, d) preserves the defect character and was selected for more detailed analysis. When extended periodically, as in Fig.~\ref{ISTW_multi_linear}(e, f), a ridge is formed which might be used to guide the direction of transported charge, as a classical nano-wire or perhaps even a quantum wire, depending on the strength of transverse electronic confinement. It is also possible to align the defects with a 60-degree bend.  Figs.~\ref{ISTW_multi_nonlin_top} and \ref{ISTW_multi_nonlin} show a sequence of such alignments that ultimately close the contour. The result is a raised {\em graphene bubble}.

\section{Conclusions}
\label{sec:conclusion}

We have shown that two fundamental defect building blocks can be used to construct a wide range of carbon nano-structures on graphene. The Inverse Stone-Thrower-Wales (ISTW) defect, recently introduced in association with graphene~\cite{LuskCarr2008a}, results from the addition of carbon ad-dimers which cause a small out-of-plane distortion in the sheet.  The Stone-Thrower-Wales (STW) defect can play a secondary role of modifying the footprint and shape of its ISTW counterpart. Both are made up of pairs of 5- and 7-membered rings.  The combination of ad-dimer addition and bond rotation allows a rich class of defect structures to be created by combining ISTW and STW defects, and these are collectively referred to as a \textit{blister taxonomy}. These blisters may be of technological importance because of their quantum confinement character and novel magnetic properties, and may also serve as sites for catalysis.  ISTW defects can be arranged linearly to form extended \textit{ridges} which may be useful in directing charge transport in graphene electronics applications.  Charge confinement may be within the ridge or between two ridges which serve as confining edges; this is the monolithic analog to graphene ribbons.  Closed ISTW contours can be used to encircle arbitrarily sized patches of pure graphene. This causes the patches to smoothly rise up from the sheet, and they are referred to as \textit{bubbles}.

To the best of our knowledge, ISTW defects have yet to be observed, although certain experiments have provided tantalizing hints. For instance, Penner {\it et al.}~\cite{heben1991} were able to produce dome-shaped mounds on HOPG using STM that were close in size to the ISTW defect.  While the dome and ISTW defect may not be the same, recent advances in scanning tunneling microscopy have made it possible to detect and characterize individual defects on graphene~\cite{stroscio2007a, stroscio2007b}. This may help to indisputably identify ISTW defects in the future. DFT calculations indicates that ISTW defects may well be found on the lip of divacancies. In fact, in graphene systems with a sufficient number of carbon adatoms, ISTW defects might also be found away from vacancies because the barrier required to form them from the chance meeting of two adatoms is relatively low.  It is therefore reasonable to search for such structures on graphene as a first step towards their experimental elucidation.  As a second step towards nanoengineering with ISTW defects, electron/ion irradiation~\cite{krasheninnikov2002, smith2001} or dicarbon molecular bombardment could be used to generate such defect structures without concern about their relative placement or type. These activities may well uncover blisters composed of combinations of STW and ISTW defects. Looking beyond isolated blisters, engineering the location of ISTW/STW defect structures will require precise atomic control. Recent technological advances allow the conclusion that the requisite fidelity is likely to be a reality in the near future. Scanning tunneling microscopy~\cite{berthe2007}, atomic force microscopy~\cite{sugimoto2005}, and eventually machines of the sort described by Allis and Drexler~\cite{allis2005}, offer the brightest prospects for such graphene nanoengineering.

The exploration and explanation of the graphene defect structures here has been undertaken primarily using DFT as an atomistically accurate method for determining energies of different structures for the purpose of evaluating stability and kinetic accessibility.  However, the general geometric features, particularly at larger scales, suggests that a more general perspective based in continuum mechanics may capture many of the essential features seen~\cite{seung1988dfm,letters2008ett}.  This is reasonable for distortions of wavelength larger than the lattice size (or bond length), given that the DFT acts to provide a classical potential energy surface for the nuclei, with harmonic constants about equilibrium.  We will consider such an approach in future work.

\section{Acknowledgements}
We also acknowledge the use computing resources provided through the Renewable Energy MRSEC program (NSF Grant No. DMR-0820518) and the Golden Energy Computing Organization (NSF Grant No. CNS-0722415) at the Colorado School of Mines.  LDC was supported by the National Science Foundation under Grant PHY-0547845 as part of the NSF Career program.  DTW was supported by the NSF (CBET-0731319, DMR-0213918) and the AFOSR (FA9550-08-1-0007).


\end{document}